# Operator-Level Description of Transient Thermal Grating Dynamics Across Transport Regimes

S. Galovic[1*], S. Jacimovski[2]

[1] *Vinca Institute of Nuclear Sciences-National Institute of the Republic of Serbia, University of Belgrade, Mike Petrovica Alasa 12-14, P.O. Box 522, 11001, Belgrade, Serbia*
[2] *University of Criminal Investigation and Police Studies, Zemun – Belgrade, Serbia, Cara Dušana 196, Zemun*

[*]correspondence: bobagal@vin.bg.ac.rs

**Keywords:** Transient thermal grating (TTG); photothermal methods; thermal transport; continuum modeling; memory effects; nonlocal temporal response; transfer function; effective transport operator; multiscale dynamics; material-property inference

## Novelty Statement

The novelty of this work lies in establishing an operator-based framework for analyzing transient thermal grating (TTG) responses and deriving explicit expressions for the complete temporal response associated with three representative transport operators: Fourier diffusion and two minimal extensions incorporating distinct forms of heat-flux memory. By retaining the full transient response, the framework provides a basis for determining thermal diffusivity and, when heat-flux memory is resolved on the relevant observation timescale and supported by the measured data, for determining the heat-propagation speed and flux-relaxation time as well. A central feature of this approach is that the identification of memory-related transport parameters does not rely solely on the presence of visible oscillations. Even when oscillatory behavior is absent from the measured space–time range, deviations from the Fourier prediction may remain encoded in the temporal shape of the TTG transient and its dependence on the grating wave number. The framework therefore makes it possible to assess whether the observed response supports a diffusive description characterized by thermal diffusivity alone or contains additional information about heat-flux dynamics. In this way, the extracted macroscopic parameters provide a physically grounded basis for the subsequent microscopic interpretation of heat transport in terms of the relevant subsystems, couplings, and relaxation processes.

## Abstract

Transient thermal grating (TTG) experiments provide a powerful means of probing material properties and heat-transport dynamics at small spatiotemporal scales through the temporal response to a spatially periodic excitation. However, quantitative interpretation of the measured transient requires a forward model that consistently connects energy deposition, subsequent heat transport, and the detected TTG observable over the experimental timescale. A unified formulation linking these stages remains insufficiently developed in the TTG context. Here, we formulate a continuum, operator-based TTG forward model that describes the sequence from energy deposition through thermal transport to the temporal evolution of the selected spatial temperature mode within a unified transfer-function framework, while leaving the instrument-specific detection process outside the scope of the present formulation. Heat-transport dynamics are represented by a temporal memory kernel, allowing unresolved relaxation processes to be incorporated into the macroscopic response. The framework provides a basis for assessing the identifiability of thermal diffusivity, heat-propagation speed, and heat-flux relaxation time when flux-memory effects are resolved on the experimental timescale. These macroscopic parameters can, in turn, inform the microscopic interpretation of the measured dynamics. Analysis of the resulting model further demonstrates that the observability of heat-flux memory is not equivalent to the presence of oscillations in the measured response. Oscillations can provide a clear signature of finite flux-relaxation dynamics, but their absence does not establish the validity of Fourier heat conduction: finite flux relaxation dynamics need not produce experimentally observable oscillations under all conditions.

## Introduction

Photothermal methods provide powerful means of determining a wide range of material properties from their response to electromagnetic excitation [1–5]. An experiment represents a cascade of interconnected processes: electromagnetic energy interacts with the material, part of the energy is absorbed and converted into excited internal degrees of freedom, the deposited energy is subsequently redistributed within the material, and the resulting dynamics is finally registered by a detection system. Thermal transport is an important part of this chain, but it represents only one of its components. To determine material properties from a measured response, a forward model that connects a physically defined excitation to the measured signal is therefore required [6–10].

At the continuum level, such forward models form the basis for the quantitative application of most photothermal methods. They provide a relation between the experimental excitation, the evolution of the relevant physical fields, and the detected response, thereby allowing effective optical, thermal, electronic, and other material properties to be inferred from measurements [11–15]. Only after such properties have been identified at the experimental level can a second question, characteristic of condensed-matter physics, be addressed: how are they related to the material structure, its microscopic degrees of freedom, and their interactions [16–19]. Effective parameters can subsequently be connected to electronic, phononic, excitonic, or other microscopic dynamics, including specific relaxation processes and couplings. The forward model therefore provides the essential link between the experiment, the determination of effective physical properties, and their subsequent microscopic interpretation.

For transient thermal grating (TTG) measurements, however, the connection between the excitation, transport dynamics, and measured response is not always formulated explicitly within a single forward framework. TTG experiments probe the temporal evolution of a spatially selected thermal mode. In the standard Fourier description, this evolution is governed by a single thermal diffusivity and follows a purely diffusive decay. However, extensive TTG studies have shown that, for a range of materials and experimental conditions, the measured transients cannot be adequately described by this Fourier evolution [20–25]. These observations have motivated the development of approaches that extend the description of transient heat conduction beyond the classical Fourier framework, including statistical and time-domain formulations [26,27].

Microscopic and kinetic approaches, including Boltzmann-transport descriptions, have long been used to interpret deviations from Fourier behavior in terms of microscopic degrees of freedom, relaxation processes, and phonon mean-free-path effects [20,21,28,29]. These approaches provide valuable insight into the microscopic origins of non-Fourier transport. However, the microscopic description of transport and the forward modelling of the experimentally measured TTG response are distinct, although closely related, problems. Several aspects of their connection require explicit consideration.

First, microscopic kinetic descriptions of TTG transport may combine material-specific information, such as *ab initio* phonon properties and scattering rates, with more general assumptions concerning the conversion of absorbed electromagnetic energy into an initial phonon distribution. Even within a phonon-gas description, a prescribed initial temperature perturbation does not uniquely determine how the deposited energy is distributed among the phonon modes. Unless this distribution is established from a model of the excitation and its coupling to the phonon system, its specification introduces an additional modelling assumption. Consequently, the material-specific character of the predicted TTG transient depends not only on the microscopic transport properties but also on how the initial state is constructed [17,30].

Second, the microscopic evolution must be distinguished from the description of transport at a specified spatial and temporal scale [30, 31]. In TTG, the grating wave number $q$ selects the spatial mode under consideration, whereas the microscopic dynamics involves a distribution of relaxation times associated with different phonon modes. Depending on the temporal scale of interest, these relaxation processes may contribute distinguishable features to the macroscopic response or combine into an effective response in which their individual contributions are not separately identifiable. Thus, the existence of a broad distribution of microscopic relaxation times does not, by itself, establish how this distribution manifests in the macroscopic transport response.

Third, microscopic detail does not necessarily translate into an unambiguous physical interpretation of the macroscopic transient. Even when material-specific phonon properties and scattering rates are obtained from *ab initio* calculations, the resulting kinetic description relies on a chosen representation of the relevant degrees of freedom, such as the phonon-gas approximation, and on assumptions concerning the preparation of the initial nonequilibrium state [28,29,32,33]. Deviations from Fourier behaviour in the resulting transient may reflect different aspects of the model, including the assumed initial phonon distribution, the relaxation dynamics of phonon modes, or correlations among modal heat fluxes. These contributions need not be uniquely distinguishable from the macroscopic transient alone.

A macroscopic transport description addresses the response at a different level: conservation laws are supplemented by constitutive relations connecting macroscopic fields and fluxes. When this constitutive relation is temporally nonlocal, its memory represents the influence of degrees of freedom that are not explicitly resolved in the macroscopic description on the temporal scale of interest, without requiring their individual microscopic contributions to be identified.

These considerations motivate a forward formulation that explicitly connects the excitation, preparation of the initial state, transport evolution, and detection within a common macroscopic framework [34]. This approach is consistent with the general continuum-transport perspective, in which conservation laws are complemented by constitutive relations specifying the material transport response [35]. The balance law alone does not determine the constitutive behavior of the heat flux; this requires a relation describing how the flux responds to the thermodynamic driving field. When this response is non-instantaneous, the constitutive relation may depend on the history of the driving field, thereby incorporating temporal memory. Non-equilibrium statistical coarse-graining provides a broader theoretical context for such memory-dependent reduced dynamics [30], while recent time-domain formulations have emphasized the role of transient transport functions and intrinsic memory across different space–time scales.

In the paper, we formulate the TTG forward problem starting from a physically defined impulsive energy excitation and explicitly distinguish the excitation, the evolution of the selected spatial mode, and the detection relation. The measured TTG signal is represented as proportional to the amplitude of this mode, with the detection mapping specified separately from the transport dynamics. The transport dynamics is described by a general causal linear operator with a temporal memory kernel. For three transport operators—Fourier diffusion and two minimal extensions incorporating distinct forms of heat-flux memory—we derive exact analytical expressions for the complete transient response in the time domain and analyse their temporal characteristics.

This analysis establishes how the temporal response can be used to identify macroscopic transport parameters, including thermal diffusivity, heat-flux relaxation time, and heat-propagation speed. Crucially, the identification of these parameters does not require oscillations in the measured transient: non-oscillatory responses can also contain identifiable signatures of heat-flux memory in their temporal structure and characteristic timescales. The relevant requirement is that these features be distinguishable within the experimental observation window. The resulting macroscopic parameters can provide a basis for subsequent microscopic interpretation in terms of material-specific degrees of freedom, their interactions, and their relaxation processes. Developing such a microscopic interpretation using appropriate microscopic, kinetic, or statistical approaches [17] is beyond the scope of the present work.

The paper is organized as follows. In Sec. II, we formulate the TTG excitation, modal description, transport operator, and detection relation, and derive the spectral form of the overall transfer function. In Sec. III, we derive the TTG transients for the considered classes of memory operators. In Sec. IV, we analyze their behavior as a function of spatial wave number and experimental temporal window, with particular emphasis on regimes in which different operators become distinguishable and on the distinction between memory dynamics and oscillatory response. In Sec. V, we discuss the physical meaning of the effective-operator parameters and their possible connection to material-specific microscopic dynamics. The main conclusions are summarized in Sec. VI.

## II. Theoretical framework

In this work, the TTG forward problem is defined as the response of the selected temperature mode $T_q(t)$ to the impulsive deposited-energy input $I_q(t)$, obtained from the coupled energy-balance and heat-flux equations. Here, $I_q(t)$ denotes the effective energy-density amplitude of the excitation, whereas $T_q(t)$ represents the resulting thermal response and its temporal evolution. The experimentally detected TTG transient is taken to be proportional to the evolution of the amplitude of this selected temperature mode [28, 29],

$$Y_{TTG}(t) = C_{det}T_q(t) \tag{1}$$

where $C_{det}$ is a proportionality factor associated with the detection process. This relation represents the detection step at the level required here and does not imply an explicit model of the complete optical detection chain.

In conventional TTG treatments, the excitation-to-temperature step is often left implicit and the evolution of $T_q(t)$ is considered directly from initial temperature state. This omission may have no apparent consequence for Fourier transport in a semi-infinite system, but it can become consequential for generalized transport, where the excitation and the transport dynamics from initial temperature state are not generally interchangeable. Here, the excitation-to-temperature step is therefore retained explicitly as part of the forward problem, while the measured TTG signal is related directly to the temporal evolution of the resulting thermal transient.

Section II develops an operator-based formulation of TTG dynamics. Starting from energy conservation and a causal linear memory constitutive relation for the heat flux, we derive a closed evolution equation for the thermally excited spatial mode. The corresponding transfer function is obtained in the Laplace domain, where the transport dynamics appears as the resolvent of an effective memory operator.

### *II.1. Optically generated heat transport: operator formulation*

We consider a photothermal system in which optical excitation generates a spatially periodic thermal perturbation. The absorbed optical energy is converted into internal energy and subsequently redistributed through the material by heat transport.

The spatial and temporal distribution of the optical excitation is described by the incident irradiance [36]:

$$I(x, z, t) = \Upsilon_0 f(t)\cos(qx)\exp(-\beta_{opt}z) \tag{2}$$

where $\Upsilon_0$ is the characteristic irradiance amplitude, $f(t)$ is a dimensionless temporal envelope, $q$ is the thermal-grating wave number, and $\beta_{opt}$ is the optical absorption coefficient both measured in [$m^{-1}$]. The coordinate $z$ measures the depth into the sample, with $z = 0$ denoting the illuminated surface, while $x$ is directed along the spatial modulation of the thermal grating. The excitation is assumed to be uniform along the *y*-direction. The cosine modulation represents the spatially periodic component of the optical interference pattern.

Assuming Beer–Lambert absorption, negligible reflection and scattering, and instantaneous conversion of the absorbed optical energy into heat [36, 37], the volumetric heat-source density is obtained from the decrease in optical irradiance along the direction of propagation:

$$Q(x,z,t) = -\frac{\partial I(x,z,t)}{\partial z} = \beta_{opt}\Upsilon_0 f(t)\cos(qx)\exp(-\beta_{opt}z) \tag{3}$$

Here, $Q(x,z,t)$ has units of [W m$^{-3}$], and the minus sign accounts for the decrease in irradiance along the direction of optical propagation.

The absorbed optical power per unit surface area is obtained by integrating the volumetric heat-source density over the sample thickness:

$$S(x,t) = \int_0^{L_z} Q(x,z,t)dz = \Upsilon_0 f(t)\cos(qx)\left(1 - \exp\left(-\beta_{opt}L_z\right)\right) \tag{4}$$

where $L_z$ is the sample thickness along the *z*-direction. The quantity $S(x,t)$ therefore provides the link between the depth-dependent optical source and the energy deposited per unit surface area. This quantity accounts for the optical energy deposited through the sample thickness (units of [Wm$^{-2}$]) without requiring the subsequent temperature field to be resolved along *z*.

In the strong-absorption limit, $\beta_{opt}L_z \gg 1$, the absorbed optical power per unit surface area approaches $S(x,t) \approx \Upsilon_0 f(t)\cos(qx)$. In this limit, the volumetric heat-source density is localized near the illuminated surface and may be approximated by

$$Q(x,z,t) \approx \Upsilon_0 \delta(z) f(t)\cos(qx) \tag{5}$$

When the optical pulse duration is also much shorter than the relevant thermal transport timescales, the temporal envelope may be approximated by an impulse. The resulting source is

$$Q(x,z,t) \approx F_0 \delta(t)\delta(z)\cos(qx). \tag{6}$$

where $F_0 = \Upsilon_0 \int_{-\infty}^{+\infty} f(t)dt$ is the incident optical fluence in [Jm$^{-2}$]. Under the assumptions of strong absorption and negligible reflection, $F_0$ equals the absorbed fluence. Accordingly, the time-integrated absorbed power satisfies $\int_{-\infty}^{+\infty} S(x,t)dt = F_0\cos(qx)$.

To construct a reduced model, we replace the depth-dependent source by an effective volumetric source distributed over the sample thickness. We define this source by $Q_{eff}(x,t) = S(x,t)/L_z$.

In the strong-absorption and impulsive-excitation limits, this gives

$$Q_{eff}(x,t) = I_q \delta(t)\cos(qx) \tag{7}$$

Here, $I_q = F_0/L_z$ is the effective excitation amplitude, with units of [Jm$^{-3}$]. This construction preserves the deposited energy per unit surface area, since

$$\int_{-\infty}^{+\infty} \int_{0}^{L_z} Q_{eff}(x,t)\, dzdt = F_0 \cos(qx)$$

In the reduced one-dimensional model, the temperature field is assumed to be uniform along the *y*- and *z*-directions and depends only on the grating coordinate *x* and time *t*. Accordingly, $T_q(t)$ represents the amplitude of the selected spatially periodic temperature mode in the reduced model; it is not obtained by solving a depth-dependent thermal problem and subsequently averaging the temperature over the sample thickness. Heat flow and heat exchange through the front and back surfaces are not resolved explicitly in this model.

The measured TTG signal can be taken as proportional to $T_q(t)$ when the detection response is linear and the signal is dominated by the selected thermal-grating mode.

Optically generated temperature perturbations in the reduced model obey local energy conservation [38–41],

$$C\dot{T}(x,t) + \frac{\partial J(x,t)}{\partial x} = I_q \delta(t) \cos(qx) \tag{8}$$

where *C* is the volumetric heat capacity [Jm$^{-3}$K$^{-1}$] and $J$ denotes the heat flux along *x* [Wm$^{-2}$].

To describe transport beyond the instantaneous Fourier relation, we introduce a general causal linear constitutive relation with temporal memory [40-43],

$$J(x,t) = -\int_0^t K(t-t') \frac{\partial T(x,t')}{\partial x} dt' \tag{9}$$

where the memory kernel characterizes the temporal response of the heat flux to the preceding thermal-gradient history. At the macroscopic level, *K*(*t*) provides an effective representation of unresolved transport dynamics [44-47]. Thus, the constitutive relation does not prescribe a particular microscopic transport mechanism; it specifies how the heat flux responds to the history of the temperature gradient at the scale of observation.

Substituting the constitutive relation into the conservation law yields

$$C\dot{T}(x,t) - \frac{\partial}{\partial x}\left[\int_0^t K(t-t') \cdot \frac{\partial T(x,t')}{\partial x} dt'\right] = I_q \delta(t) \cos(qx) \tag{10}$$

The resulting equation is linear and time-invariant under the assumptions of small temperature excursions, constant material properties, and a memory kernel that depends only on the time difference $t - t'$.

This equation (Eq. 10) is therefore the closed macroscopic evolution equation for optically generated perturbation of the spatially reduced thermal field.

Because the TTG excitation is periodic along the $x$-direction and uniform along the $y$-direction, the reduced temperature field can be represented by the excited spatial Fourier mode, $T(x,t) = T_q(t)\cos(qx)$. Here, $T_q(t)$ is the effective amplitude of the selected TTG mode in the reduced model. Substitution into Eq. (10), followed by extraction of the common spatial factor $\cos(qx)$, yields:

$$C\dot{T}_q(t) + q^2 \left[ \int_0^t K(t-t') \cdot T_q(t') dt' \right] = I_q \delta(t) \tag{11}$$

Equation (11) governs the effective temporal response of the selected TTG mode, with the unresolved temporal transport dynamics encoded in the memory kernel.

### *II. 2. Transfer function formulation*

For an impulsive temporal excitation, the determination of $T_q(t)$ can also be formulated as an evolution problem starting from the state produced by the impulse. In this representation the initial temperature is $T_q(0^+) = I_q/C$. However, for general memory-dependent constitutive laws, this temperature alone does not fully specify the subsequent evolution; additional information required to close the evolution problem must be obtained from, or specified consistently with, the constitutive relation for the heat flux. Depending on the memory kernel, the resulting formulation may involve additional dynamical variables or higher-order evolution equations.

In the present work, we instead adopt the Laplace-domain approach, which provides a direct representation of the excitation–response relation and allows different constitutive laws to be treated within a common transfer-function framework.

The linear, time-invariant structure of Eq. (11) permits the TTG response to be represented in the Laplace domain by a transfer function, or resolvent, of the effective thermal transport operator. Introducing the Laplace transform, $\tilde{f}(s) = \int_0^\infty e^{-st} f(t) dt$, in Eq. (11) and assuming zero initial temperature perturbation and zero heat flux before the pulse, Eq. (11) becomes

$$CsT_q(s) + q^2 \widetilde{K}(s) T_q(s) = I_q \tag{12}$$

The temperature response can therefore be written as

$$T_q(s) = H(q,s) I_q \tag{13}$$

where

$$H(q,s) = \frac{1}{Cs + q^2\widetilde{K}(s)} \quad (14)$$

is the transfer function, or resolvent, of the effective inhomogeneous thermal transport operator.

The transfer function maps the imposed modal excitation onto the corresponding temperature response, with its temporal structure determined by the memory kernel. For the impulsive excitation considered here, the resulting transient is directly the Green response of the transport operator.

For a specified excitation and wavevector $q$, the analytical structure of the thermal response is consequently determined by the Laplace-domain memory kernel. Different kernels represent different realizations of the same generalized transport framework, with diffusive, relaxing, and propagating responses appearing as limiting cases [38,40,43].

In the short-memory limit $K(t) = k\delta(t)$, the transport becomes Markovian and the Fourier diffusion equation is recovered [38,48,49]:

$$\widetilde{K}(s) \approx k \quad (15)$$

where $k$ is the thermal conductivity [$Wm^{-1}K^{-1}$], α=k/C is the thermal diffusivity [$m^2s^{-1}$] and $\delta(t)$ is Dirac $\delta$-function.

An exponentially decaying memory kernel, $K(t) = ({}^{k}/_{\tau})e^{-\frac{t}{\tau}}$, yields the hyperbolic Cattaneo model [38, 40, 41, 50,51] ,

$$\widetilde{K}(s) \approx \frac{k}{1+s\tau} \quad (16)$$

A step-function memory kernel, $K(t) = Cc_{eff}^2 h(t)$, gives the corresponding undamped hyperbolic limit [38 ,43,52],

$$\widetilde{K}(s) \approx \frac{Cc_{eff}^2}{s} \quad (17)$$

where $h(t)$ is Heaviside step function.

Thus, diffusive, damped wave-like, and undamped wave-like responses emerge from the same effective transport framework and are distinguished by the temporal structure of the memory kernel.

These three cases are representative rather than exhaustive. The same operator framework can accommodate other memory kernels [53-56], including those leading to anomalous subdiffusive [55] or fractional wave-like thermal responses [57-60]. Such memory structures provide an effective representation of unresolved multiscale dynamics without requiring its microscopic origin to be specified at the forward-model level.

The resulting formulation provides a continuum forward model for TTG dynamics in which the measured transient is determined by the optical excitation, the selected grating wavevector $q$, and the effective transport operator. This provides a common basis for comparing different transport dynamics under the same excitation and experimental conditions, and for assessing which features of the response become distinguishable within a given temporal observation scale.

## III. TIME-DOMAIN RESPONSE

The TTG transient is obtained by inverse Laplace transformation of the transfer function [61-63],

$$T_q(t) = I_q \mathcal{L}^{-1}\{H(q,s)\}(t) \tag{18}$$

whose zero–pole structure determines the temporal morphology of the response. Since the transfer function is itself determined by the Laplace-domain memory kernel, different memory kernels give rise to distinct time-domain response classes.

In the following, we derive the time-domain responses for representative memory kernels by inverse Laplace transformation. The resulting response classes follow from the corresponding analytic structure of the resolvent.

### Short memory kernel

For a memoryless (Markovian) transport process, the kernel reduces to the constant limit given by Eq. (15). The corresponding transfer function becomes

$$H(q,s) = \frac{1}{C}\frac{1}{s+\alpha q^2} \tag{19}$$

The inverse Laplace transform yields a single exponential relaxation:

$$T_q^{(0)}(t) = \frac{I_q}{C}\exp(-\alpha q^2 t) \tag{20}$$

Since symbol $I_q$ denotes the effective excitation amplitude, with units of [$\mathrm{Jm^{-3}}$] and $C$ is the volumetric heat capacity, $I_q/C$ has units of temperature and determines the initial amplitude of the excited TTG temperature mode.

### Finite (fading)-memory (Cattaneo) kernel

For a finite flux-relaxation time, the memory kernel takes the form given by Eq. (16), and the corresponding transfer function is

$$H(q,s) = \frac{1}{C}\frac{\left(s+\frac{1}{\tau}\right)}{s^2+\left(\frac{1}{\tau}\right)s+\left(\frac{1}{\tau}\right)\alpha q^2} \tag{21}$$

The response is determined by the resulting pole–zero structure. The zero is located at $s = -1/\tau$ while the poles are given by the roots of the characteristic equation

$$s^2+\left(\frac{1}{\tau}\right)s+\left(\frac{1}{\tau}\right)\alpha q^2 = 0 \tag{22}$$

The discriminant

$$\Delta = 1 - 4\alpha q^2 \tau \tag{23}$$

determines a critical wavevector $q_c$,

$$q_c = \frac{1}{2\sqrt{\alpha\tau}} \tag{24}$$

which separates three qualitatively different dynamical regimes.

For $q < q_c$ the inverse Laplace transform yields a sum of exponential terms:

$$T_q^{(\tau)}(t) = \frac{I_q}{C}\left[A_1(q,\tau)exp(\lambda_1 t) + A_2(q,\tau)exp(\ \lambda_2 t)\right] \tag{25}$$

where

$$\lambda_{1/2} = -\frac{1}{2\tau} \pm \frac{1}{2\tau}\sqrt{1-4\alpha q^2\tau} \tag{26}$$

$$A_1(q,\tau) = \frac{\left(\frac{1}{\tau}+\lambda_1\right)}{\lambda_1-\lambda_2} \tag{27}$$

$$A_2(q,\tau) = -\frac{\left(\frac{1}{\tau}+\lambda_2\right)}{\lambda_1-\lambda_2} \tag{28}$$

At the critical wavevector $q = q_c$, the two poles coalesce, and the inverse Laplace transform yields

$$T_q^{(\tau)}(t) = \frac{I_q}{C}exp\left(-\frac{t}{2\tau}\right)\left[1+\frac{t}{2\tau}\right] \tag{29}$$

with

$$\lambda_1 = \lambda_2 = -\frac{1}{2\tau} \tag{30}$$

For $q > q_c$, the poles form a complex-conjugate pair, and the inverse Laplace transform yields a damped oscillatory response:

$$T_q^{(\tau)}(t) = \frac{I_q}{C} \mathrm{A}(\mathrm{q}, \tau) exp\left(-\frac{t}{2\tau}\right) sin(bt + \phi(q, \tau)) \quad (31)$$

where

$$b = \frac{\sqrt{4\alpha q^2 \tau - 1}}{2\tau} \quad (32)$$

$$\mathrm{A}(\mathrm{q}, \tau) = \frac{\sqrt{\frac{1}{4\tau^2} + b^2}}{b} \quad (33)$$

$$\phi = arctg(2b\tau) \quad (34)$$

For the finite-memory (Cattaneo) kernel, the temporal response is determined by the pole structure of the resolvent and changes qualitatively with the grating wavevector relative to the critical value $q_c$. Specifically $q < q_c$ gives two distinct real poles and therefore a bi-exponential relaxation; at $q = q_c$, the poles coalesce, producing the critical response; and for $q > q_c$ the poles form a complex-conjugate pair, resulting in damped oscillations. Thus, the transition from non-oscillatory to oscillatory TTG response is a direct consequence of the pole structure of the same finite-memory transport operator.

**Long-memory kernel**

For the long-memory reference case, the Laplace-domain kernel is taken in the singular form given by Eq. (17), giving the transfer function

$$H(q, s) = \frac{1}{C} \frac{s}{s^2 + c_{eff}^2 q^2} \quad (35)$$

The inverse Laplace transform then gives the corresponding undamped oscillatory response:

$$T_q^{(\infty)}(t) = \frac{I_q}{C} cos(c_{eff} q t) \quad (36)$$

In the singular long-memory limit, the poles lie on the imaginary axis, yielding an undamped oscillatory response. For comparison with the finite-τ Cattaneo model, this limit corresponds to $\tau \to \infty$ with finite $k/\tau = C c_{eff}^2$. This limiting case therefore represents a qualitatively distinct dynamical regime from the damped oscillations produced by finite flux-relaxation time.

The time-domain responses derived above follow directly from the analytic structure of the resolvent, which is determined by the chosen Laplace-domain memory kernel under the assumptions of linearity and time-translation invariance. The resulting zero pole structure thus provides a direct link between the temporal form of the TTG transient and the effective memory encoded in the transport operator.

## IV. Scaling Structure of TTG Dynamics

The memory term in macroscopic heat-flux constitutive models can be understood, in the spirit of the Mori–Zwanzig (MZ) framework, as representing the influence of degrees of freedom that are not explicitly retained in the reduced description but contribute to the temporal correlations governing transport. From this perspective, the form of the memory kernel reflects the effective temporal structure of these unresolved dynamics. Different idealized kernels can therefore emphasize different characteristic features of heat transport, depending on the relation between the relaxation times of the unresolved degrees of freedom and the timescale on which the transport response is examined.

The delta-kernel and step-kernel models represent two distinct idealizations of the constitutive response. Within the interpretation adopted here, the delta kernel describes the limit in which the relevant memory effects are effectively instantaneous relative to the characteristic observation timescale. It yields a diffusive macroscopic description characterized by thermal diffusivity, while propagative features associated with finite-speed heat-flux dynamics are not explicitly retained. The step kernel, in contrast, defines a limiting hyperbolic description characterized by a finite heat-propagation speed, with dissipative attenuation neglected in this model. In the present formulation, these two limits are interpreted, respectively, as descriptions appropriate when memory effects are effectively unresolved on the observation timescale and when the constitutive dynamics are represented by the idealized propagation-dominated limit. This interpretation is specific to the models considered here and does not imply a universal correspondence between these kernels and the microscopic relaxation times of all underlying degrees of freedom.

The finite-memory formulation retains the temporal structure of the constitutive response rather than imposing either of these limiting descriptions in advance. It therefore allows us to examine how the transient response relates to the diffusive and hyperbolic reference models as the relative timescales change. In this analysis, the characteristic observation timescale, denoted by $\tau_{obs}$, is the physical timescale relative to which the response is analyzed. It may correspond, for example, to millisecond, microsecond, nanosecond, or shorter dynamics. It is distinct from the duration of a plotted time interval and from the temporal sampling interval. Changing $\tau_{obs}$ does not alter a fixed constitutive kernel; rather, it changes the relative scales against which the finite-memory response is assessed. Moreover, changing the observation timescale or temporal resolution does not alter the microscopic dynamics of the material. Rather, it changes which portions of the dynamics can be resolved in the measured response and, consequently, which features can be represented by an effective macroscopic transport description.

To characterize the coupled temporal and spatial dependencies of the transient thermal-grating (TTG) response, we introduce two dimensionless parameters: : $\gamma = \tau/\tau_{obs}$ and $\beta = 1/(\alpha q^2 \tau_{obs})$. where $\tau$ is the flux-relaxation time, $\alpha$ is the thermal diffusivity, and $q$ is the wavevector of the spatially periodic excitation. The parameter $\gamma$ compares the intrinsic relaxation time with the observation timescale, whereas $\beta$ compares the diffusive timescale associated with the excited mode, $1/(\alpha q^2)$ , with the same observation timescale. Equivalently, $\beta$ relates the diffusion length over $\tau_{obs}$ to the spatial scale selected by the TTG excitation. Both parameters participate in the modal analysis, although the criticality condition depends on a combination of them rather than on either parameter independently. Retaining them separately allows us to examine their respective contributions to the transient response while accounting for the selected spatial mode. The modes are classified as subcritical, critical, or supercritical according to the relation between $q$ and the critical wavevector $q_c$.

The following analysis examines this scaling structure and its implications for interpreting TTG transients. Table 1 summarizes the principal dimensional quantities and dimensionless parameters used in the formulation, together with their definitions, units, and physical interpretation.

All graphical representations show the normalized transient response, with the temperature-mode amplitude scaled by the characteristic temperature $I_q/C$ and time expressed in terms of the dimensionless variable $t/\tau_{obs}$. This normalization facilitates comparison of the transient responses and makes their short-time and long-time characteristics more readily distinguishable relative to the prescribed observation timescale.

**Table 1. Principal dimensional and dimensionless parameters of the TTG transport model.**

| Symbol | Definition | Units | Physical meaning |
|---|---|---|---|
| $T_q$ | TTG temperature mod | K | Temperature amplitude of the selected TTG spatial mode |
| $q$ | | $m^{-1}$ | TTG grating wavevector |
| $C$ | | $J\ m^{-3}K^{-1}$ | Volumetric heat capacity |
| $k$ | | $W\ m^{-1}\ K^{-1}$ | Thermal conductivity |
| $\alpha = k/C$ | | $m^2\ s^{-1}$ | Thermal diffusivity |
| $\tau$ | | s | Flux-relaxation (memory) time |
| $I_q$ | | $J\ m^{-3}$ | Amplitude of impulsive TTG heat-flux excitation |
| $q_c$ | | $m^{-1}$ | Critical wavevector |
| $c_{eff}$ | | $m\ s^{-1}$ | Heat propagation speed |
| $\tau_{obs}$ | | s | Experimental observation time scale |
| $\gamma$ | $\gamma = \tau/\tau_{obs}$ | | Memory time relative to observation time |
| $\beta$ | $\beta = 1/(\alpha q^2 \tau_{obs})$ | | Diffusive modal time scale relative to observation time |

## *IV.1. Memory kernels and unresolved dynamics at the experimental observation scale*

We first examine the two asymptotic memory limits of the TTG response: the memoryless and long-memory limits, focusing on their transient dynamics and modal scaling. We then consider the finite-memory response, which exhibits distinct dynamical regimes depending on the spatial mode, and examine how its temporal evolution relates to the two asymptotic limits as the dimensionless parameter $\gamma = \tau/\tau_{obs}$ varies.

IV.1.1 Asymptotic memory limits

The two singular limits (Eqs. 20, 36) provide asymptotic reference responses for examining the role of the TTG wavevector. In the memoryless limit, variation of $q$ changes the relaxation rate while preserving the monotonic character of the response. In the long-memory limit, variation of $q$ changes the propagation frequency and therefore the temporal scale of the oscillatory response.

These two limits establish a fundamental distinction between $q$-dependent relaxation and $q$-dependent propagation. The corresponding modal responses are illustrated in Fig. 1.

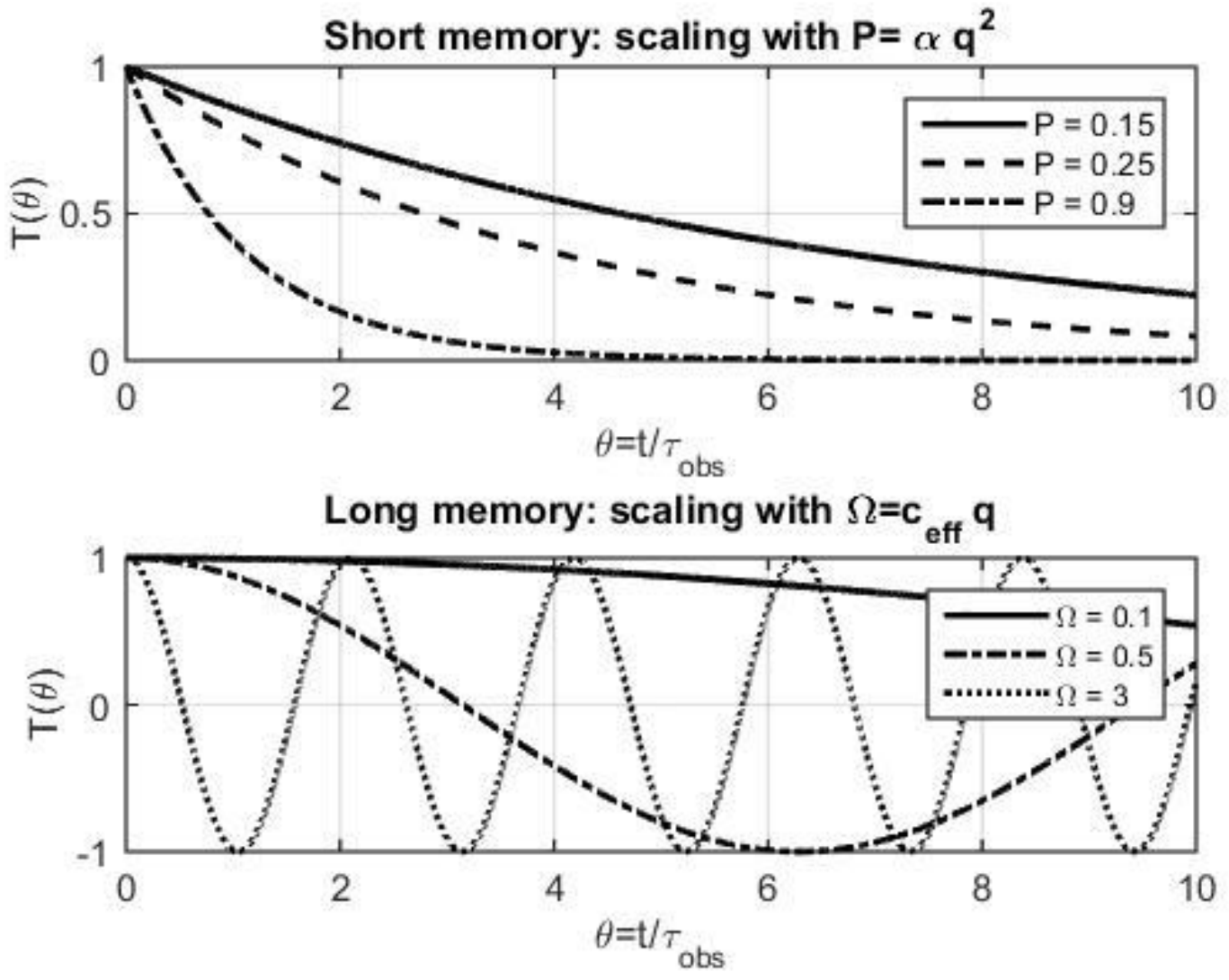

**Figure 1. Asymptotic modal responses. (a) Short-memory limit (Eq. 20), showing exponential relaxation of the selected TTG mode. The decay rate is determined by the thermal diffusivity and the grating wavevector. (b) Long-memory limit (Eq. 36), showing oscillatory modal evolution. The oscillation frequency is determined by the effective propagation velocity and the grating wavevector. The two limits provide asymptotic reference responses for the finite-memory dynamics.**

In the short-memory limit, the characteristic decay rate scales as $\alpha q^2$. Thus, changing the grating wavevector changes the temporal relaxation rate while preserving the monotonic character of the response. This $q^2$ scaling provides the characteristic signature of diffusive relaxation and defines the effective diffusivity that can be extracted from the modal decay.

In the long-memory limit, the modal response is propagating, with a characteristic angular frequency that scales linearly with $q$. Thus, changing the grating wavevector does not merely rescale a relaxation rate; it changes the temporal frequency of the propagating mode. The corresponding linear q-scaling provides the characteristic signature of propagation and allows the effective propagation velocity to be identified from the modal dynamics.

The observability of this propagation signature, however, depends on the temporal scale of the measurement. Since the oscillation period is

$$t_q = \frac{2\pi}{\omega_q}$$

changing $q$ changes the temporal scale that must be resolved. For sufficiently small $q$, the oscillation period may exceed the available observation window, so that only a fraction of the propagating evolution is observed. For sufficiently large $q$, the oscillation period becomes shorter, and individual oscillations may become unresolved by finite temporal sampling or may be compressed into the early-time part of the transient.

Importantly, the absence of visibly resolved oscillations does not imply the absence of the underlying propagating dynamics. The asymptotic limits therefore provide two distinct scaling references: $q^2$ scaling of the diffusive relaxation rate and linear $q$ scaling of the propagation frequency. The finite-memory response considered below connects these limits and determines how the corresponding signatures are transformed, attenuated, or rendered unresolved within a finite experimental observation window.

IV.1.2 Finite memory kernel

For the finite-memory kernel, the temporal evolution depends simultaneously on the relaxation time $\tau$, the thermal diffusivity $\alpha$, and the selected spatial mode $q$. The resulting modal response is organized by the critical wavevector $q_c$ (Eq. 24), which separates relaxational and oscillatory regimes.

The finite-memory response is examined separately for the critical, subcritical, and supercritical modes. The three regimes correspond directly to the pole classification established in Sec. III: $q < q_c$ gives two real poles, $q = q_c$ a repeated pole, and $q > q_c$ a complex-conjugate pole pair.

The limiting values of $\gamma$ provide asymptotic reference cases for the dependence of the observed response on the experimental timescale. In the limit $\gamma \to 0$, the flux-relaxation time is much shorter than the observation timescale. The flux memory therefore relaxes too rapidly to be resolved experimentally, and the constitutive response approaches its memoryless Fourier limit. Consequently, the measured transient appears diffusive on the observation timescale. In the opposite limit, $\gamma \to \infty$, the flux-relaxation time is much longer than the observation timescale. The flux relaxation is then effectively frozen during observation, and the response approaches the long-memory hyperbolic limit. These limits do not represent different microscopic transport mechanisms, but rather distinct asymptotic regimes of the same memory-dependent dynamics as observed on different temporal scales. The parameter $\gamma$ characterizes the relation between the flux-relaxation time and the observation timescale, while the selected spatial mode determines how the transport dynamics is manifested in the TTG response.

Critical mode $q = q_c$

At the critical wavevector $q = q_c$, the two poles of the transfer function coalesce and the response is described by Eq. (29). The resolvent therefore exhibits a degenerate pole structure, and the corresponding time-domain response remains monotonic.

Figure 2 illustrates the dependence of the response on the dimensionless parameter $\gamma$.

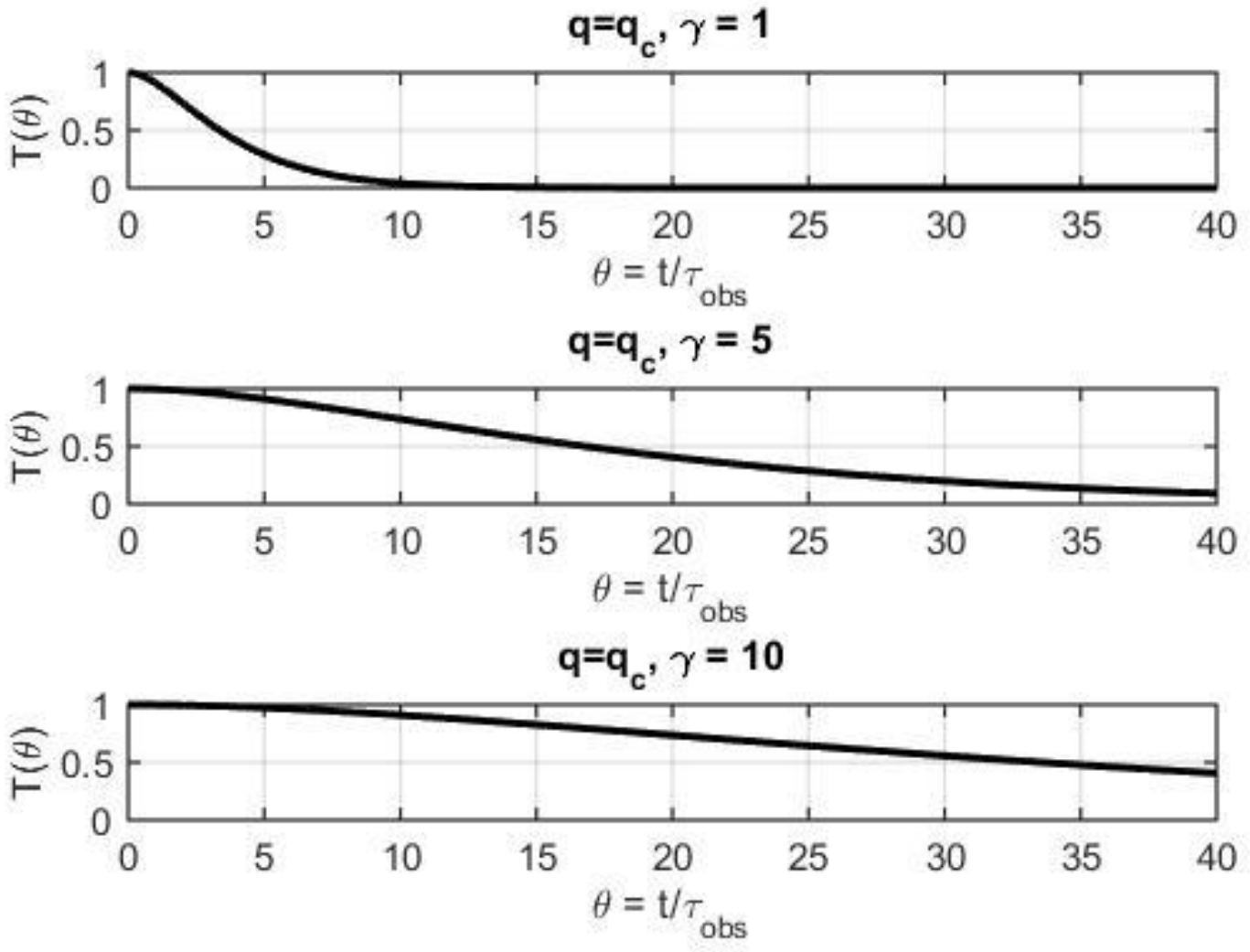

**Figure 2. Finite-memory response at the critical wavevector $q = q_c$. Temporal evolution of the TTG mode for several values of $\gamma$. Increasing $\gamma$ extends the characteristic relaxation over a longer interval on the observation timescale and shifts the change in curvature towards later times.**

The critical-mode response provides a complementary perspective on the temporal signatures of finite-memory transport. Its defining feature is the evolution of its curvature with the dimensionless parameter $\gamma$. For small $\gamma$, the concave portion becomes barely discernible, and the response decreases towards zero along an almost entirely convex curve. At intermediate values of $\gamma$, the transition from concave to convex curvature is clearly visible at intermediate times. As $\gamma$ increases further, the interval of concave behavior becomes more extended, while a convex portion with a slower decay remains present at later times (long-time response). At the critical wavevector, the normalized response has the same functional form when expressed in terms of $t/\tau$. Since the response is plotted against $t/\tau_{obs}$, changing $\gamma = \tau/\tau_{obs}$ changes the temporal scale of the response relative to the observation timescale, without changing its intrinsic functional form.

Overall, for small values of $\gamma$, the characteristic relaxation and the change in concavity occur within a relatively short interval after excitation. As $\gamma$ increases, the non-exponential character of the response remains observable beyond the short-time response.

Subcritical mode $q < q_c$.

For wavevectors below the critical value, the resolvent has two distinct real poles, and the response is therefore represented by two exponential relaxation modes (Eq. 25). The TTG transient is consequently a superposition of fast and slow relaxation contributions, with their relative amplitudes determined by the relaxation time $\tau$ and the selected spatial mode q.

Figure 3 illustrates this behavior as a function of γ at fixed $\beta$. For small γ, the finite-memory response remains close to the diffusive reference over the observation window, while its long-time behavior is dominated by the slowest pole. As γ increases, the contribution of the fast pole becomes progressively more pronounced, producing a distinct short-time departure from both the purely diffusive response and the asymptotic slow relaxation. At longer times, the fast contribution decays more rapidly and the response approaches the single-exponential behavior associated with the slowest pole. Thus, finite memory may first appear as a short-time deviation from diffusive relaxation while leaving the long-time response apparently single-exponential. The relative visibility of the two relaxation scales therefore depends on the observation window as well as on the intrinsic parameters of the transport operator.

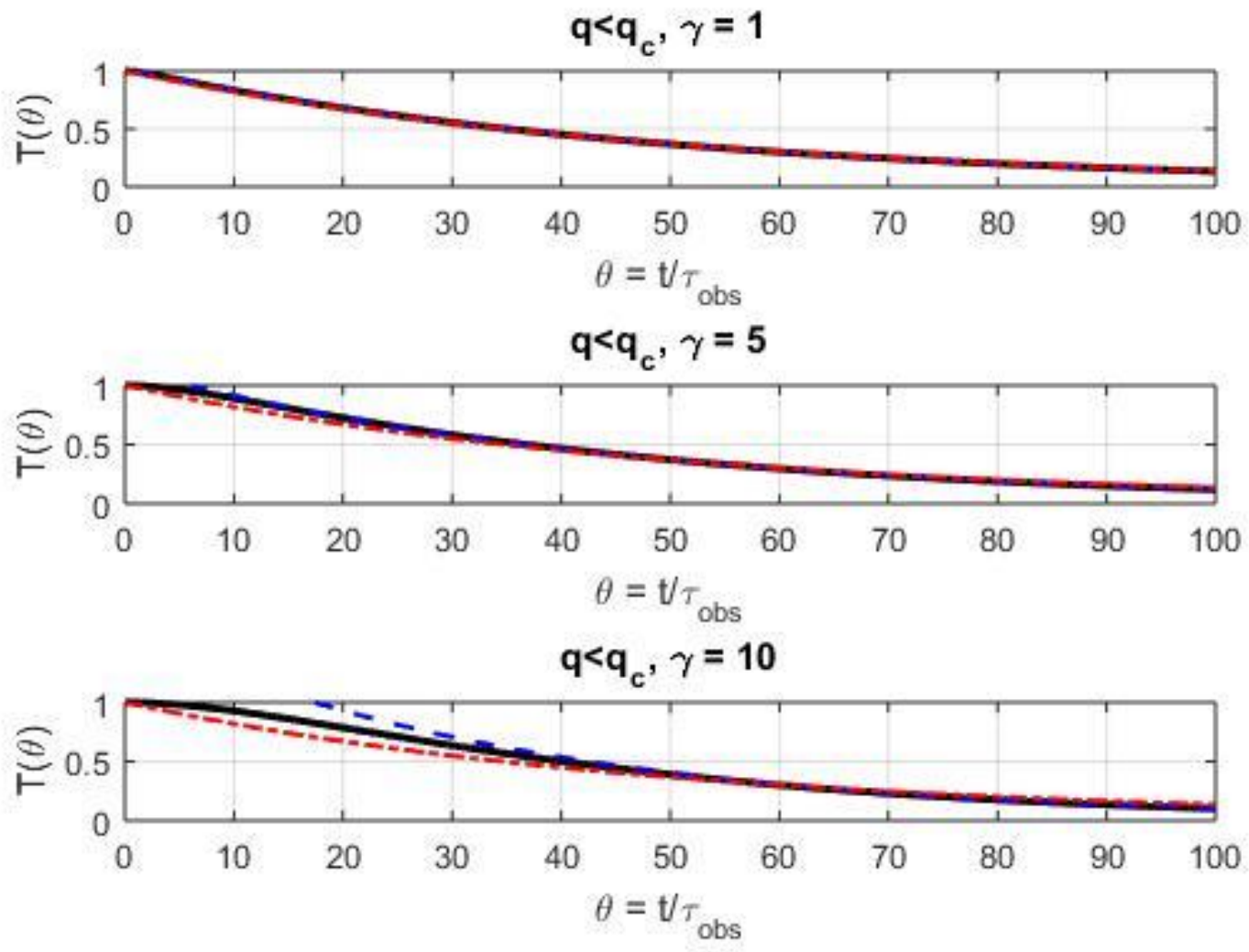

**Figure 3. Subcritical finite-memory dynamics $q < q_c$. Temporal evolution of the TTG mode for several values of $\gamma$ at fixed $\beta$=50. The solid curves show the exact finite-memory response. For each value of $\gamma$, the dashed curve represents the contribution of the slowest relaxation mode, while the dash-dot curve denotes the diffusive reference response. Increasing $\gamma$ progressively reveals the short-time transient associated with the fast relaxation mode, while the long-time response becomes increasingly governed by the slowest relaxation contribution.**

Supercritical mode $q > q_c$

For wavevectors above the critical value, the two poles become a complex-conjugate pair and the response assumes the form of a damped oscillation (Eq. 31). Figure 4 illustrates the dependence of the response on $\gamma$. For small values of $\gamma$, only a limited portion of the oscillatory structure is resolved within the observation window. As $\gamma$ increases, the oscillatory contribution becomes increasingly pronounced and a larger fraction of the underlying wave-like dynamics is resolved. Over the observation window, the finite-memory response consequently approaches the undamped long-memory reference as the effect of finite relaxation becomes progressively less prominent.

The oscillatory response remains bounded by an exponentially decaying envelope determined by the finite relaxation time τ. Thus, propagation and dissipation are simultaneously encoded in the complex-conjugate pole structure, while the visibility of the oscillations depends on their characteristic frequency, the relaxation time, and the experimental observation window.

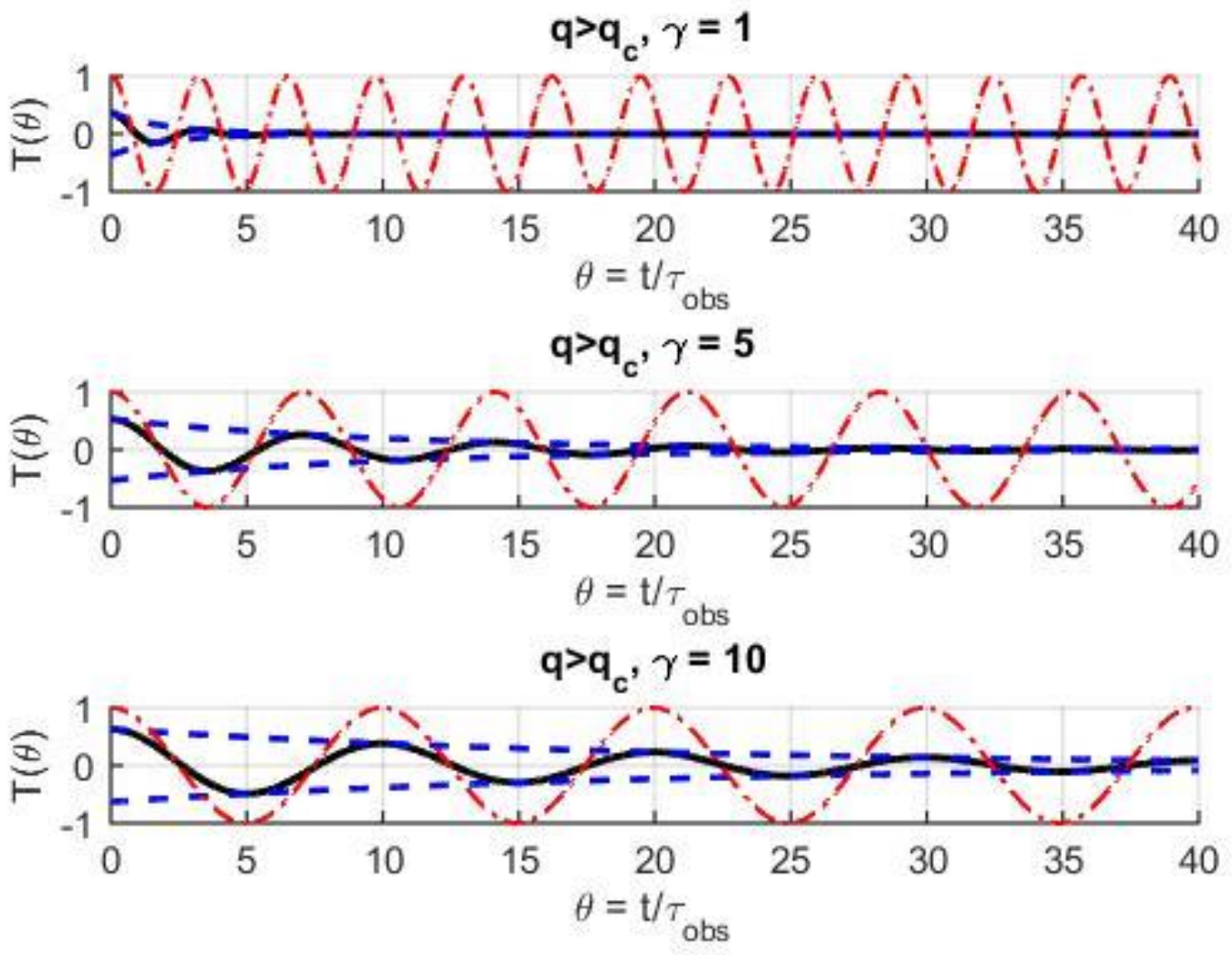

**Figure 4. Supercritical finite-memory dynamics $q > q_c$. Temporal evolution of the TTG mode for several values of $\gamma$ at fixed $\beta$=0.25. The solid curve shows the exact finite-memory response. The dash-dot curve represents the oscillatory long-memory reference, while the dashed curve denotes the corresponding exponential envelope. Increasing γ enhances the visibility of oscillatory features and progressively brings the response closer to the long-memory limit. The oscillatory component remains bounded by an exponentially decaying envelope determined by the finite relaxation time.**

The preceding results show that a monotonic TTG transient does not by itself establish the absence of unresolved transport dynamics. In the memoryless limit, the $q$-dependence is captured by the diffusive scaling of a single effective diffusivity, whereas finite memory introduces an additional relaxation timescale and modifies the temporal morphology of the response. At fixed $q/q_c$, the parameter $\gamma = \tau/\tau_{obs}$, controls whether this additional dynamics is experimentally resolved beyond short time response; once resolved, the relation between $q$ and $q_c$ determines whether the response is subcritical, critical, or supercritical. We therefore now fix γ and vary β to examine how different TTG modes are manifested within the same temporal observation window.

### *IV.2. Finite-Memory Dynamics across TTG Modes*

For a fixed memory time relative to the observation window, different values of β, corresponding to different TTG wavevectors, produce different temporal responses. Since $\beta = 1/(\alpha q^2 \tau_{obs})$, variation of β at fixed α and $\tau_{obs}$ directly corresponds to variation of the selected spatial mode $q$, with decreasing β corresponding to increasing $q$. The effect of $q$ is not restricted to a rescaling of the response time, as in the asymptotic limits. In the finite-memory regime, changing $q$ modifies the pole structure of the resolvent and can move the response between relaxational, critical, and oscillatory regimes. The $q$-dependence therefore determines not only the characteristic timescale of the TTG transient but also its temporal morphology within a given experimental observation window. The grating wavevector thus provides an experimental control parameter for probing different spectral regimes of the same transport operator.

The following analysis examines this dependence at fixed γ, thereby keeping the temporal resolvability of the memory process fixed while varying the selected TTG mode**.** Once the finite-memory dynamics are resolved within the observation window, variation of $q$ probes how these dynamics are manifested across different spatial modes.

Subcritical modes $q < q_c$

For subcritical modes, increasing $q$ corresponds to decreasing β and therefore to faster modal evolution within the fixed observation window. The relative temporal placement of the fast and slow relaxation contributions consequently changes as the selected TTG mode is varied. As shown in Fig. 5, decreasing β progressively compresses the finite-memory transient into an earlier portion of the observation interval. The fast relaxation contribution therefore becomes increasingly confined to early times, making the transient-to-asymptotic crossover more difficult to resolve experimentally.

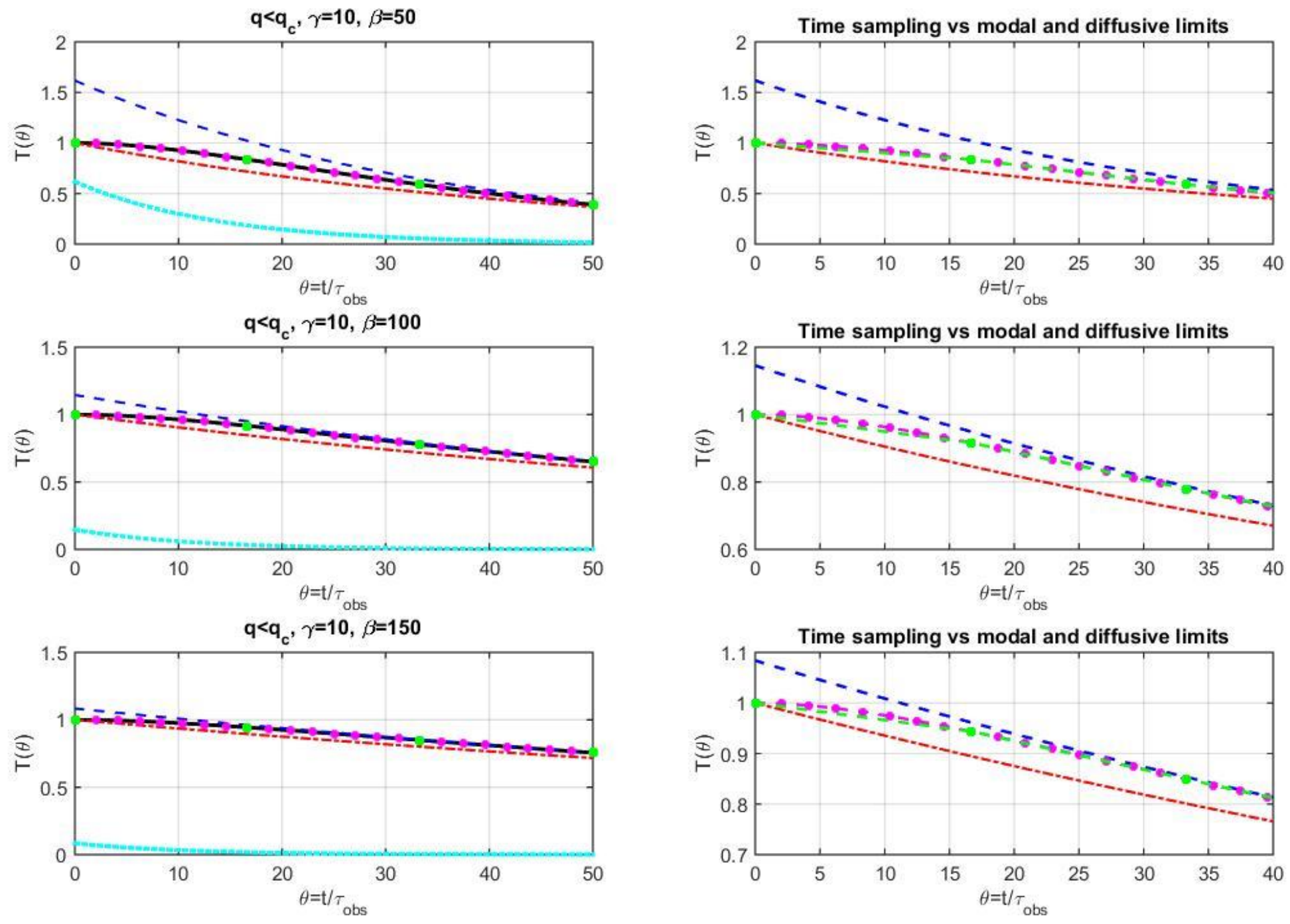

**Figure 5. Observability of finite-memory signatures in the subcritical regime ($q < q_c$) for fixed γ and different values of β. Left panels show the exact finite-memory response (black solid curve), its decomposition into the slow relaxation mode (blue dashed curve) and the fast transient contribution (cyan dashed curve), and the diffusive reference (red dash-dot curve). Dense and sparse temporal sampling are represented by magenta and green markers, respectively. Right panels show the corresponding sampled responses together with the slow relaxation mode and the diffusive reference. Decreasing β, corresponding to increasing q, compresses the fast finite-memory transient toward earlier times. Consequently, the transient-to-asymptotic crossover may become experimentally unresolved even though the response remains governed by the finite-memory operator.**

The exact response is a superposition of fast and slow relaxation modes. For larger β, corresponding to smaller $q$, their crossover extends over a larger fraction of the observation window and is therefore more readily resolved. As β decreases, the crossover is progressively compressed toward earlier times. Finite temporal sampling further reduces the observability of this local signature. Sparse sampling may therefore obscure the curvature transition associated with the fast relaxation process. Importantly, this loss of a visible transient signature does not imply the disappearance of memory. Once the fast contribution has decayed, the response remains governed by the finite-memory operator and approaches its slow pole rather than the Fourier solution.

Thus, Fig. 5 illustrates the distinction between the visibility of a memory signature and the presence of memory itself. The transient crossover may become experimentally unresolved as the modal timescale changes, while the underlying finite-memory dynamics remain encoded in the slow-pole response.

Supercritical modes ($q > q_c$)
For supercritical modes, the finite-memory response is governed by a complex-conjugate pole pair and therefore contains a damped oscillatory component. At fixed γ, decreasing β corresponds to increasing $q$ and hence to a shorter characteristic modal timescale. The resulting oscillatory structure is progressively compressed within the fixed observation window. As shown in Fig. 6, decreasing β shifts the oscillatory extrema toward shorter times, making individual oscillations progressively more difficult to resolve.

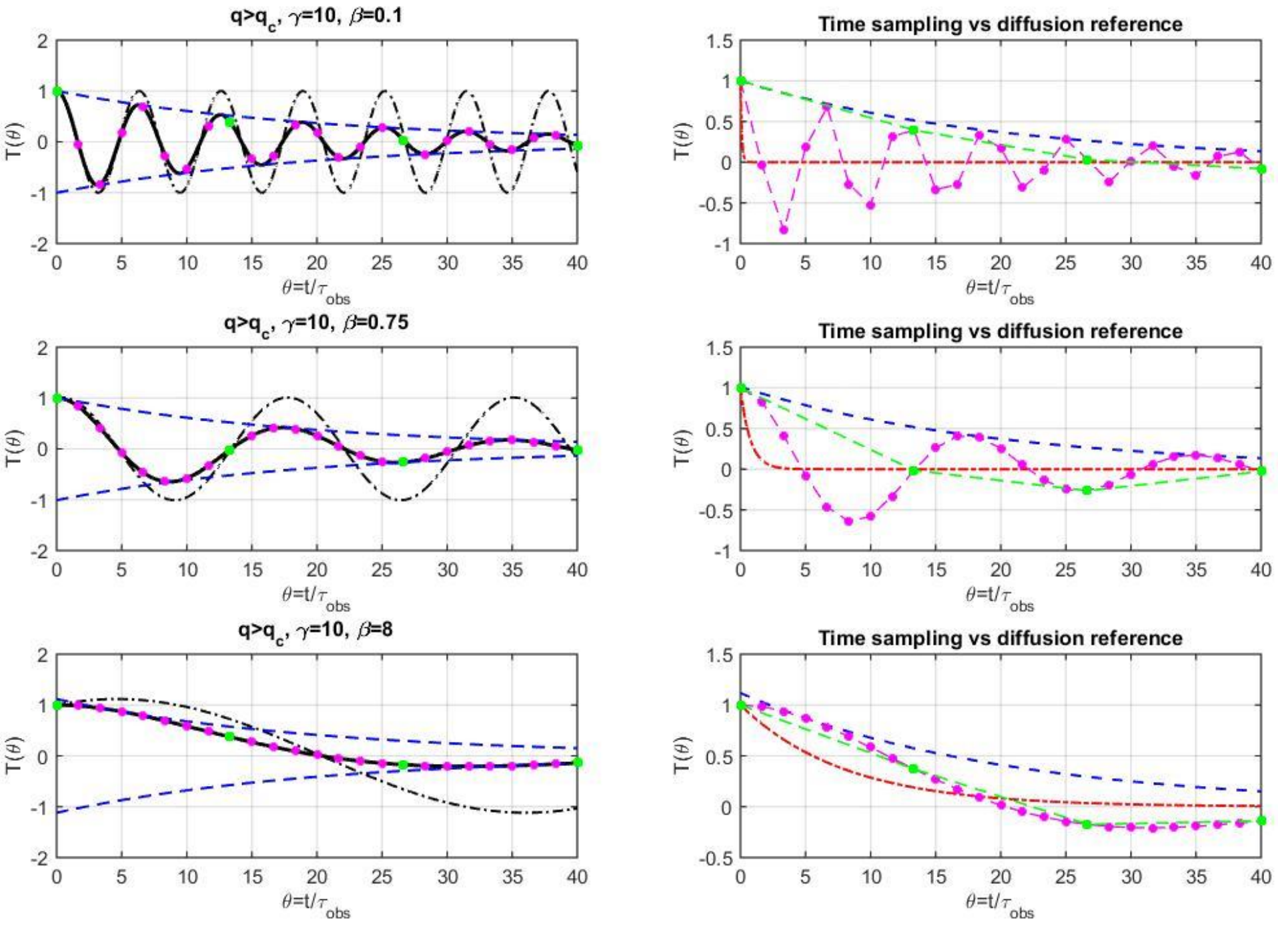

**Figure 6. Observability of finite-memory signatures in the supercritical regime ($q > q_c$) for fixed γ and different values of β. The black solid curve denotes the exact finite-memory response, the blue dashed curve the corresponding damping envelope, and the black dash-dot curve the undamped oscillatory reference. Magenta and green markers represent dense and sparse temporal sampling, respectively. Right panels compare the sampled responses and the corresponding envelope with the diffusive reference (red dash-dot curve). Decreasing β, corresponding to increasing q, compresses the oscillatory structure in time and may render**

**individual extrema experimentally unresolved. The underlying complex-pole dynamics, however, remain present. Thus, loss of visible oscillations does not imply a transition to memoryless diffusion.**

Finite temporal sampling can further suppress these features, so that the measured transient may appear smooth or nearly monotonic even though the underlying response remains governed by the complex-conjugate pole pair. Thus, oscillation visibility is not equivalent to memory visibility. Resolved oscillations provide a clear signature of the propagating branch of the finite-memory operator, but their absence does not establish memoryless diffusion. Other signatures of finite memory may remain in the damping envelope and in the deviation of the transient from the diffusive response. The oscillatory response remains bounded by an exponentially decaying envelope determined by the finite relaxation time τ. Thus, propagation and relaxation are simultaneously encoded in the complex-pole structure of the resolvent.

The visibility of the oscillatory response is therefore controlled not only by the existence of complex poles but also by the relation between their characteristic frequency, the relaxation time, and the experimental observation window.

Taken together, Figs. 5 and 6 demonstrate that the visibility of memory effects should not be identified with the visibility of a particular temporal morphology. In the subcritical regime, the fast memory-related transient may become compressed beyond experimental resolution, whereas in the supercritical regime the oscillatory extrema may become unresolved. In both cases, the underlying finite-memory dynamics remain encoded in the response of the same transport operator.

Consequently, the $q$-dependence provides information that cannot be inferred from the morphology of a single transient. Changing $q$ changes the spectral response of the same operator and can move characteristic memory signatures into or out of the experimentally accessible temporal range. Loss of visible oscillations should therefore be interpreted as loss of oscillation visibility, not necessarily as loss of memory.

This also shows why a single-exponential fit of a TTG transient cannot, in general, be interpreted as a measurement of an effective diffusivity. For a genuinely memoryless operator, the decay rate follows the $q^2$ scaling of the diffusive limit and uniquely defines the diffusivity. For a finite-memory operator, however, the apparent decay rate depends on the spectral regime and on the portion of the transient selected for fitting. A fit to a subcritical relaxation branch, the critical response, or the damping envelope of a supercritical response does not represent the same physical quantity and need not define a unique $\alpha_{eff}$.

In the supercritical regime, fitting the damping envelope primarily probes the relaxation time associated with the finite-memory operator rather than a Fourier diffusivity.

## V. Discussion

The results presented above show that the interpretation of TTG transients depends not only on the transport model adopted, but also on the relation between the intrinsic dynamics of the effective transport operator and the space–time scale of the measurement. In the present formulation, the measured transient is treated as the response of a continuum transport operator to a physically defined impulsive excitation. The memory kernel provides an effective representation of transport dynamics that are not explicitly resolved at the observational scale, while the grating wavevector determines which spatial mode of the operator is probed. This separates the experimentally controlled observation scale from the microscopic origin that may ultimately underlie the effective transport parameters.

A central consequence is that finite transport memory cannot, in general, be identified with an oscillatory TTG response. For the finite-memory operator considered here, the same underlying transport description produces distinct temporal regimes depending on the selected wavevector. Subcritical modes are governed by two real poles and exhibit a superposition of fast and slow relaxation processes, while the critical mode corresponds to coalescence of the two poles and remains monotonic. Only in the supercritical regime do the poles become complex conjugates and generate damped oscillations. Thus, oscillations constitute a clear signature of the propagating branch of the finite-memory response when they are experimentally resolved, but their absence does not establish a memoryless Fourier regime. Memory may instead remain observable through non-exponential relaxation or systematic deviations from the diffusive response.

The dependence on the grating wavevector provides an additional distinction between asymptotic and finite-memory transport. In the memoryless limit, variation of $q$ changes the modal decay rate according to the diffusive $q^2$ scaling. In the long-memory limit, it changes the propagation frequency approximately linearly with $q$. For finite memory, however, $q$ also modifies the pole structure of the resolvent and can therefore move the response between relaxational, critical, and oscillatory regimes. The grating wavevector consequently acts not only as a probe of a characteristic transport length scale, but also as an experimental control parameter for accessing different spectral regimes of the same effective transport operator.

The observation scales introduces a further level of interpretation. As demonstrated in Figs. 5 and 6, changing $q$ at fixed observation time changes the temporal placement of the characteristic features of the response. In the subcritical regime, the finite memory-related contribution may be compressed into the earliest part of the transient and become unresolved, leaving an apparently simple long-time relaxation. In the supercritical regime, the oscillatory extrema may likewise become compressed beyond the available temporal sampling. In both cases, the underlying finite-memory dynamics remain present even when their most conspicuous temporal signature is no longer resolved. Consequently, experimental visibility of a particular feature should not be equated with the existence or absence of the corresponding dynamical process.

This observation has direct implications for the extraction of effective thermal properties from TTG measurements. A single-exponential fit has an unambiguous interpretation as a thermal diffusivity only when the response is genuinely governed by the memoryless diffusive operator and the corresponding $q^2$ scaling is observed. For a finite-memory response, an apparent decay rate depends on the spectral regime and on the temporal portion of the transient selected for fitting. In the subcritical regime, a fit may predominantly capture the slow pole after the fast contribution has decayed; at the critical point, the response is intrinsically non-exponential; and in the supercritical regime, fitting the damping envelope primarily probes the relaxation scale of the finite-memory operator rather than a Fourier diffusivity. Therefore, an apparent decay rate obtained from a single transient need not represent a unique material property unless its $q$-dependence is consistent with the memoryless scaling.

It is important to note that even when this scaling is observed, distinguishing a genuinely diffusive response from a finite-memory response whose characteristic features remain unresolved requires sufficient experimental information and consideration of measurement noise, temporal resolution, and instrumental effects.

The present formulation is deliberately not intended to assign a particular microscopic mechanism to the effective memory kernel [64,65]. More detailed material-specific approaches may explicitly resolve selected microscopic degrees of freedom and their interactions and thereby relate the observed TTG response to particular microscopic mechanisms [66-72]. Such approaches are essential when the microscopic origin of the observed transport is the question of interest. The role of the present continuum formulation is complementary: it provides an intermediate forward level at which the experimentally resolved dynamics can first be characterized in terms of effective operator parameters, independently of a particular microscopic model or microscopic ansatz. The subsequent connection of these parameters to material-specific degrees of freedom and interaction mechanisms can then be formulated as a separate interpretive problem.

## VI. Conclusions

We have formulated a continuum, operator-based forward model for transient thermal grating (TTG) dynamics that connects energy deposition, the evolution of the selected spatial mode, and detection within a unified transfer-function framework. Starting from a physically defined impulsive energy deposition, heat transport is described by a causal linear operator with a temporal memory kernel. The resulting TTG response is governed by the resolvent of this operator, whose zero–pole structure characterizes the temporal dynamics of the measured mode.

The analysis demonstrates that the observable consequences of finite heat-flux memory are not restricted to oscillatory responses. For the finite-memory operator considered here, subcritical modes exhibit two relaxation scales, the critical mode remains monotonic while displaying a non-exponential temporal evolution, and supercritical modes exhibit damped oscillations associated with complex-conjugate poles. Oscillations therefore provide a clear signature of the propagating branch of finite-memory dynamics, but their absence does not establish the Fourier limit.

We further show that the observability of these signatures depends on the spatial and temporal scales probed by the experiment. Whether finite relaxation dynamics can be resolved depends on the relation between the memory time and the experimental observation timescale, while the grating wave number determines the spectral regime of the selected mode. Changing the grating wave number can therefore shift characteristic memory-related features into or out of the experimentally accessible temporal range without changing the underlying material dynamics. The disappearance of a visible oscillatory or fast-transient feature may also reflect limited temporal resolution rather than the absence of heat-flux memory.

The analysis also establishes the theoretical conditions for determining transport parameters from TTG transients. When heat-flux memory dynamics are resolved, the response provides a basis for determining the thermal diffusivity, the heat-flux relaxation time, and the heat-propagation speed, including in the non-oscillatory regime. When memory-related features are not experimentally resolved, a Fourier description may provide an adequate approximation. However, the absence of resolved memory signatures does not by itself establish the validity of Fourier diffusion. This approximation must be assessed against the available data, including the dependence on the grating wavevector and the measurement resolution. Interpreting a single-exponential decay rate as thermal diffusivity requires consistency with the diffusive $q^2$ scaling. When finite-memory dynamics are experimentally resolved, the temporal response and its dependence on the grating wavevector must instead be analyzed within a memory-dependent transport model. The practical determination of thermal diffusivity, heat propagation speed, and heat flux relaxation time, as well as the distinction between resolved and unresolved memory dynamics, depends on the characteristics of the detection system and the available experimental information.

The proposed framework provides a macroscopic forward level for TTG experiments in which transport dynamics and their identifiable parameters can be inferred on the space–time scales resolved by the experiment, without prescribing a microscopic transport mechanism *a priori*. The resulting effective operator parameters can subsequently constrain the microscopic interpretation of heat transport in terms of material-specific degrees of freedom, their interactions, and their relaxation processes. This separation between experimentally resolved transport dynamics and their subsequent microscopic interpretation provides a general framework for analyzing TTG transients beyond the classical Fourier description of heat transport.

**Data availability statement:** The data that supports the findings of this study are available from the corresponding author upon reasonable request.

**Competing interests:** The author declares no competing interests.

**Founding:** This research was funded by the Ministry of Science, Technological Development and Innovations of the Republic of Serbia (Contract No. 451-03-33/2026-03/ 200017).

**Acknowledgement:** The authors gratefully acknowledge Prof. Natalie Germann for the kind invitation to contribute to the special collection *Tribute to John C. Slattery*. The operator-level formulation presented in this work emphasizes the distinction between the conservation law and the constitutive relation governing the material's transport response, while allowing the latter to incorporate temporal memory. This perspective provides a natural connection to the general framework of transport phenomena developed by Slattery.